\documentstyle[12pt,epsfig]{article}
\newcommand{\be}{\begin{equation}}
\newcommand{\ee}{\end{equation}}
\newcommand{\bea}{\begin{eqnarray}}
\newcommand{\eea}{\end{eqnarray}}
\newcommand{\beas}{\begin{eqnarray*}}
\newcommand{\eeas}{\end{eqnarray*}}
\newcommand{\bd}{\begin{displaymath}}
\newcommand{\ed}{\end{displaymath}}
\def\shiftleft#1{#1\llap{#1\hskip 0.04em}}
\def\shiftdown#1{#1\llap{\lower.04ex\hbox{#1}}}
\def\thick#1{\shiftdown{\shiftleft{#1}}}
\def\b#1{\thick{\hbox{$#1$}}}

\begin{document}

\baselineskip 0.8cm

\centerline{\Large\bf{Nucleon deformation and atomic 
spectroscopy}~\footnote{published in Can. J. Phys. {\bf 83}, 455 (2005).} }
\vskip 0.6 cm
\centerline{\large A.J. Buchmann}
\vskip 0.2 cm
\begin{center}
Institute for Theoretical Physics, University of T\"ubingen  \\
Auf der Morgenstelle 14, D-72076 T\"ubingen, Germany \\
e-mail:alfons.buchmann@uni-tuebingen.de  \\
\end{center}

\begin{abstract}
Recent ineleastic electron-proton scattering experiments
have led to rather accurate values for the  $N \to \Delta$ transition
quadrupole moment $Q_{N \to \Delta}$.
The experimental results imply a prolate (cigar-shaped) intrinsic 
deformation of the nucleon. The nonsphericity of the proton's charge 
distribution might be seen in the spectrum of atomic hydrogen.
The possibilities and limitations for 
determining the geometric shape of the nucleon in an atomic physics experiment 
are  discussed.
\end{abstract}

\section{Introduction}


Elastic electron-proton scattering experiments 
and atomic spectroscopy measurements have shown that the positive charge 
of the proton is distributed over a finite volume, and that the 
charge radius of the proton is about $r_p= 0.9$ fm~\cite{Kar99}.
Furthermore, the proton's radial charge density $\rho(r)$ could be determined
from the Fourier transform of the charge form factor measured in elastic
electron-proton scattering. However, the proton charge distribution most
certainly has an angular dependence, i.e., $\rho({\bf r})=\rho(r,\Theta)$.
This angular dependence of the proton charge density is not directly 
accessible in elastic scattering.


Inelastic photon and electron scattering experiments provide further details
about the inner structure of the nucleon. In particular
they show that the proton has a spectrum of excited states.
When photons of sufficient energy impinge on a hydrogen target, 
the proton maybe excited to its lowest lying excited state with spin 3/2, 
called the $\Delta^+(1232)$, where the  number in parentheses is 
the energy (mass) of this state in units of MeV. 
The $\Delta^+$ resonance decays quickly via the strong interaction into the 
proton ground state of 939 MeV and the lightest strongly interacting 
particle, the pion, with a mass of about 140 MeV.  
The lifetime of the $\Delta^+$ is of the order of $10^{-23}$ s, i.e., 
the time it takes for a gluon to travel across the 
proton radius of 10$^{-15}$ m. This corresponds to a line width of about
100 MeV.

From a multipole analysis of the electromagnetic $p \to \Delta^+$ 
transition amplitude
we have infered that the proton charge density deviates 
from spherical symmetry. 
In fact, recent experimental determinations of the $p \to \Delta^+$
transition quadrupole moment~\cite{Ber03,Tia03,Bla01} provide 
evidence for a prolate (cigar-shaped) intrinsic 
deformation of the proton~\cite{Hen01}.
 
\section{Structure of the nucleon}

\subsection{Quark model}

The structure of the nucleon~\footnote{ 
The proton and neutron, which have nearly the same mass of about 939 MeV
can (with respect to strong interactions) be regarded as one particle, 
called nucleon $N(939)$, that exists in two charge states. 
Similarly, the first excited state of the nucleon, 
the $\Delta(1232)$ exists in four charge states $\Delta^-,\Delta^0, \Delta^+, 
\Delta^{++}$, which are energetically almost degenerate, i.e., all
four states have a mass of about 1232 MeV. Mathematically, 
this symmetry is decribed by the SU(2) isospin  
group, whose generators are, in analogy to the spin, the Pauli matrices. 
The $N$ has isospin 1/2 and the $\Delta$ has isospin 3/2.
The quarks building up the $N$ and $\Delta$ 
have approximately the same mass and thus have isospin 1/2.} 
ground state and its 
excitation spectrum is explained in the quark model~\cite{Fra74}. 
In the quark  model the proton and neutron consist 
of three spin 1/2 quarks which come in two charge states, 
called up and down quarks, i.e., $p(uud)$ and $n(ddu)$. 
Here, $u$ and $d$ denote the up-quark with electric charge $e_u=2/3$ 
and down-quark with charge $e_d=-1/3$.  
The first excited states of the proton and neutron, 
the  $\Delta^+(uud)$  and $\Delta^0(ddu)$ 
with spin $J=3/2$, have the same quark content as the $p$ and $n$ with spin 
$J=1/2$. Thus we expect the properties of the $\Delta$ to be closely related
to those of the $N$ ground state. Formally, $N$ and $\Delta$ properties are 
related because there exists a higher symmetry than isospin, called SU(6) 
spin-flavor symmetry, which combines the $N$ isospin doublet with $J=1/2$, 
and the $\Delta$ isospin quartet with $J=3/2$ together with other 
states in the same 56-dimensional spin-flavor supermultiplet.

 
\begin{figure}[tbh]
\label{splitting}
\centering{
\includegraphics{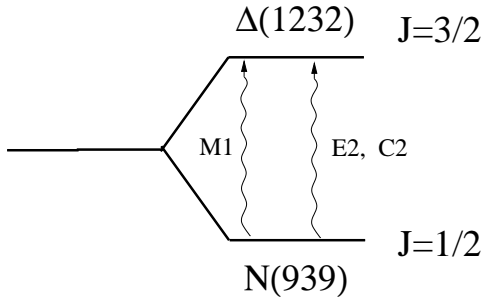}}
\caption{\label{figure:splitting} 
The hyperfine splitting between the $N(939)$ and $\Delta(1232)$ 
due to the spin-dependent interaction between quarks. 
This splitting is the nucleon physics
analogue of the 21 cm line between the hyperfine states $F=0$ and $F=1$ in
 atomic hydrogen.
In nucleon spectroscopy, 
an excited state is usually denoted by the name of
the resonance followed by its energy [MeV]  
in parentheses.}
\end{figure}

The strong interaction between the quarks is modelled as 
a long range harmonic oscillator confinement force, 
which guarantees that the colored quarks cannot 
be observed as free particles. In order to explain the spectrum in greater
detail, for example, the mass splitting between the $N(939)$ with 
spin 1/2 and the 
$\Delta(1232)$ with spin 3/2 (see Fig.~\ref{figure:splitting}), 
the long-range spin-independent confinement forces must be supplemented 
by short-range spin-dependent terms.
Spin-dependent forces arise, e.g., from one-gluon exchange between quarks. 
The one-gluon exchange potential (OGEP) is analogous to the Fermi-Breit 
interaction between the electron and proton in atomic hydrogen, 
except for a factor $\b{\lambda}_i \cdot 
\b{\lambda}_j$, which reflects that quarks carry color 
charges, and the replacement of the photon-electron coupling constant 
$\alpha$ by the strong gluon-quark coupling $\alpha_s$. The most important
terms of the one-gluon exchange potential (spin-orbit terms are omitted) are 
\bea
\label{gluon}
V^{OGEP} ({\bf r}_i,{\bf r}_j)\! \!\! \! \!&  = & 
\! \! \!\! \!{\alpha_{s}\over 4} 
\b{\lambda}_i\cdot\b{\lambda}_j \!\Biggl \lbrace
{1\over r}- \!  
{\pi\over m_q^2} \left ( 1+{2\over 3}
\b{\sigma}_{i}\cdot\b{\sigma}_{j} \right ) \delta({\bf r}) 
- \!  {1\over 4m_q^2} {1\over r^3}
\left ( 3\b{\sigma}_i\cdot  {\hat {\bf r} }  
\b{\sigma}_j\cdot{\hat{\bf r}}\! 
-  \! \b{\sigma}_i\cdot\b{\sigma}_j \right ) \Biggr \rbrace .
%
\eea
Here, ${\bf r}={\bf r}_i-{\bf r}_j$ is the distance between the i-th 
and j-th quark; $ {\hat {\bf r}}= {\bf r}/r$ is the corresponding unit vector,
$\b{\sigma}_i$ is the spin operator 
(Pauli matrices), and $\b{\lambda}_i$ is the color operator 
(Gell-Mann matrices). The constituent quark mass of the up and down quarks 
is denoted as $m_q$. Usually one chooses 
$m_q=M_N/3$ where $M_N$ is the nucleon mass. 
In the entire paper we use natural units $\hbar=c=1$.

The $1/r$ term is called color-Coulomb interaction, and the
$\b{\sigma}_i \cdot \b{\sigma}_j$ term is refered to as color-magnetic
interaction. The matrix element of the operator 
$\b{\lambda}_i\cdot\b{\lambda}_j$ between color singlet
states, such as the $N$ and $\Delta$, is readily evaluated and gives 
a factor -8/3. Therefore, the color magnetic interaction
is repulsive between quark pairs coupled to spin 1 and attractive
between quark pairs coupled to spin 0.  

The mass difference between the $N$ and $\Delta$ masses of 293 MeV 
is almost completely determined by the color-magnetic interaction 
in Eq.(\ref{gluon}); the $N-\Delta$ mass difference is often refered 
to as hyperfine splitting (see Fig.~\ref{figure:splitting})
in analogy to the $10^{-6}$ eV splitting 
between the $F=0$ and $F=1$ hyperfine states in the hydrogen atom ground 
state. The huge difference of fourteen orders
of magnitude between the hyperfine splitting in the nucleon and hydrogen atom
is mainly due to the different size of both 
systems\footnote{
The matrix element of the $\delta$-function in Eq.(\ref{gluon}) is
proportional to $1/R^3$, where $R$ is a measure of the spatial extension 
of the system, i.e., the quark core radius $b \approx 0.5 $ fm in the case 
of the nucleon,  and the Bohr radius $a_0 \approx 0.5$ \AA \  in the case 
of the hydrogen atom.}. 


\subsection{The electromagnetic $N \to \Delta$ transition}
\label{sec:NDelta}

\begin{figure}[tbh]
\centering{
\includegraphics{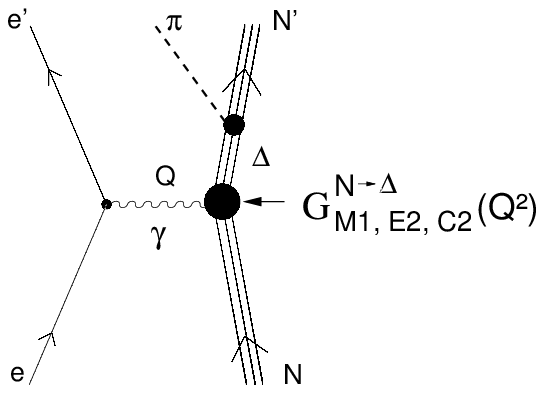}}
\caption{\label{figure:scattering} 
The excitation of the $\Delta(1232)$ resonance by a virtual 
photon $\gamma$ of momentum $Q$ is described 
by the three electromagnetic transition form factors 
$G^{N \to \Delta}_{M1}(Q^2)$, $G^{N \to \Delta}_{E2}(Q^2)$, 
and $G^{N \to \Delta}_{C2}(Q^2)$. 
Their contribution to the inelastic electron-nucleon cross section 
can be obtained by analyzing the angular distribution of the decay pions in 
coincidence with the scattered electron. In the limit $Q\to 0$, the cross 
section is described by the magnetic dipole moment $\mu_{N \to \Delta}$ 
and the charge quadrupole moment $Q_{N \to \Delta}$. }
\end{figure}

If the $\Delta$ resonance is produced in an electromagnetic process 
(see Fig.~\ref{figure:scattering}), parity invariance
and angular momentum conservation restrict the $N \to \Delta$ excitation to
magnetic dipole (M1), electric quadrupole (E2), and charge or (Coulomb) 
quadrupole (C2) transitions. Although the E2 and C2 amplitudes are small 
(about 1/40 of the dominant magnetic dipole transition amplitude) they 
are nevertheless important because their nonzeroness provides evidence 
for a deviation of the nucleon charge distribution from spherical 
symmetry~\cite{Ber03,Tia03,Bla01,Hen01}. In the quark model,
the $\Delta(1232)$ is obtained from the $N(939)$ ground state either
by flipping the spin of a single quark (M1 transition) or by flipping 
the spins of two quarks (E2 or C2 transition)~\cite{Buc97}. In the case of
the quadrupole transitions there is also
a small contribution coming from e.g., lifting a single quark from an 
$S$ state in the nucleon into an orbitally excited $D$ state in the $\Delta$. 

With the quark model it is also possible to calculate how much the proton 
charge distribution deviates from spherical symmetry and which degrees of 
freedom are responsible for this deformation~\cite{Hen01}. 
A multipole expansion of the nucleon charge density operator $\rho$ 
up to quadrupole terms leads to the following invariants in spin-isospin
space
\begin{eqnarray}
\label{structure}
\rho & = & A \sum_i^{3} e_i -B \sum_{i < j}^{3}
 e_i \biggl [  2\, \b{\sigma}_i \cdot \b{\sigma}_j
-( 3 \b{\sigma}_{i\, z} \, \b{\sigma}_{j \, z} 
-\b{\sigma}_i \cdot \b{\sigma}_j ) \biggr ], 
\end{eqnarray}
where $\b{\sigma}_{i \, z}$ is the $z$-component of the Pauli spin
matrix of quark $i$, and $e_i=\frac{1}{6}(1 + \b{\tau}_{i \, z})$ 
is the quark charge where $\b{\tau}_{i \, z}$ is the third component
of the Pauli isospin matrix. 
The constants $A$ and $B$ in front of the one- and two-quark terms
parametrize the orbital and color matrix elements
so that $\rho$ is only an operator in spin-isospin space.
It is important to note that the one-body operator (the A term) 
in Eq.(\ref{structure}) represents the valence quark degrees of freedom, 
whereas the two-body operator (the B term)  even though it also acts 
on quark variables provides an effective description of the 
quark-antiquark degrees of freedom in the physical $N$ 
and $\Delta$~\cite{Hen01,Buc91,Buc97}. 
Three-quark currents are omitted here, but were taken 
into account in Ref.~\cite{Hes02}.  

After evaluating Eq.(\ref{structure}) between quark model spin-isospin 
wave functions
the following relations between the $p$, $n$, and  $\Delta^+$
charge radii, and between the $p\to \Delta^+$ transition quadrupole moment 
$Q_{p \to \Delta^+}$ and the neutron charge radius $r^2_n$ 
were derived~\cite{Buc97}
\be
\label{rnquad}
r^2_p -r^2_{\Delta^+} = r^2_n, \qquad 
 Q_{p \to \Delta^+}= \frac{1}{\sqrt{2}} \, r^2_{n},  
\ee
where $r_p^2$ and $r_{\Delta^+}^2$ are the proton and $\Delta^+$ charge radii.
Inserting the experimental neutron charge radius~\cite{Kop95} 
$r^2_n=-0.113(3)$ fm$^2$ in the second relation of Eq.(\ref{rnquad}) 
we obtain $Q_{p \to \Delta^+}=-0.08$ fm$^2$. This agrees well
with recent determinations of $Q_{p \to \Delta^+}$ from inelastic 
electron-proton scattering data which yield
$Q_{p \to \Delta^+}({\rm exp})=-0.0846(33)$ fm$^2$~\cite{Tia03},
and $Q_{p \to \Delta^+}({\rm exp})=-0.108(9)$ fm$^2$~\cite{Bla01}.
The second relation is the zero momentum transfer limit of a more 
general relation between the $N \to \Delta$ charge 
quadrupole transition form factor  $G^{N\to \Delta}_{C2}(Q^2)$ and the 
elastic neutron charge form factor $G_C^n(Q^2)$
\bea
\label{ffrel2}
G_{C2}^{N \to \Delta}(Q^2) & = &  -\frac{3\,\sqrt{2}}{Q^2}\,  G_C^n(Q^2), 
\eea
which is experimentally satisfied for a wide range of 
momentum transfers~\cite{Buc04}.

\begin{figure}[tbh]
\centering{
\includegraphics{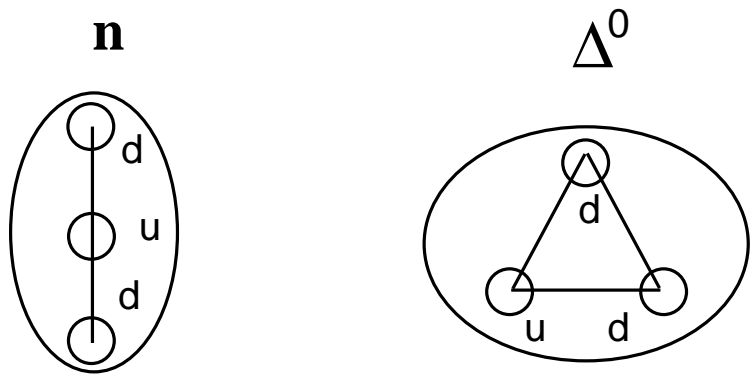}}
\caption{\label{figure:shapes} 
Qualitative picture of the neutron (left) and $\Delta^0$ (right) 
charge distributions in the quark model. Although 
the deformation is depicted here as residing in the valence quark 
distribution, in reality a spherical quark core is surrounded by 
a deformed cloud of quarks and antiquarks~\cite{Hen01}. A similar picture
is obtained for the $p$ and $\Delta^+$ by interchanging $u$ and $d$ quarks. 
}
\end{figure}

These relations are a consequence of the underlying SU(6)
spin-flavor symmetry  and its breaking by the spin-dependent two-body terms 
in Eq.(\ref{structure}). As we have noted before, the spin-spin interaction 
in Eq.(\ref{gluon}) is repulsive between quark pairs in a spin 1 state
and attractive in quark pairs with spin 0. This explains
why the $\Delta^+$, which contains only quark pairs coupled to spin 1 
is heavier than the proton. Similarly, the electromagnetic counterparts 
of these spin-dependent terms in Eq.(\ref{structure}) explain 
why $r^2_{\Delta^+} > r^2_p$, why the neutron charge radius is negative, 
and why the neutron has a prolate intrinsic deformation.  
This can be qualitatively understood as follows. The two 
down quarks in the neutron are always in a spin 1 
state\footnote{
In order to satisfy the Pauli principle the 
total three-quark wave function, which is the direct product
of orbital, spin-isospin, and color wave functions, 
must be completely antisymmetric under permutation of any two quarks. 
Because the color singlet
$N$ and $\Delta$ wave functions are 
antisymmetric in color space, the product of orbital and spin-isospin
wave functions must be symmetric. For orbitally symmetric
ground state S-wave functions this in turn means that the spin-isospin 
wave functions must be completely symmetric. The latter 
can only be achieved if the two down quarks in the neutron, which are 
necessarily in an isospin 1 state, are simultaneously in a spin 1 state.}. 
Consequently, the spin-spin force pushes them further apart than 
an up-down quark pair. This results in an elongated 
(prolate) charge distribution with the up quark in the middle,
and at the same time in a negative neutron charge radius. 
On the other hand, in the neutral $\Delta$, where all quark pairs are
in a spin 1 state, there is an equal distance between up-down and 
down-down quark pairs. This corresponds to an equilateral triangle (oblate) 
configuration of the charges and a zero charge radius of the 
neutral $\Delta$ (see Fig.~\ref{figure:shapes}).


\section{Shape of the nucleon}

\subsection{Intrinsic quadrupole moment of the nucleon}

In order to learn something about the shape of a spatially extended 
particle one has to determine its {\it intrinsic} quadrupole 
moment. The {\it intrinsic} quadrupole moment of a nucleus
\be 
Q_0=\int \! \!d^3r \, \rho({\bf r}) \,  (3 z^2 - r^2), 
\ee
is defined with respect to the body-fixed frame. 
If the charge density is concentrated along the $z$-direction 
(symmetry axis of the particle),  
the term proportional to $3z^2$ dominates, $Q_0$ 
is positive, and the particle is prolate (cigar-shaped).
If the charge density is concentrated in the equatorial plane perpendicular
to $z$, the term proportional to $r^2$ prevails, $Q_0$
is negative, and the particle is oblate (pancake-shaped).

The intrinsic quadrupole moment $Q_0$ must be distinguished
from the {\it spectroscopic} quadrupole moment $Q$ measured in 
the laboratory frame.
In the collective nuclear model \cite{Boh75}, the relation between the 
observable spectroscopic quadrupole moment $Q$ and the intrinsic quadrupole 
moment $Q_0$ is 
\begin{equation}
\label{collective}
Q= {3 K^2 -I\, (I+1) \over (I+1) (2I+3) } \, Q_0,
\end{equation}
where $I$ is the total spin of the nucleus,
and $K$ is the projection of $I$ onto the $z'$-axis in the body fixed frame
(symmetry axis of the nucleus)~\footnote{
In the following the nuclear spin
is denoted as {\bf I} and the total angular momentum of the electrons
is denoted by ${\bf J}$.}.
The factor between $Q$ and $Q_0$ is due to the quantum mechanical 
precession of the deformed charge distribution with body-fixed
symmetry axis $z'$ around the lab frame $z$-axis along ${\bf I}_z$.
Therefore, in the lab frame one does not measure the intrinsic
quadrupole moment directly but only its projection onto the lab frame
quantization axis. As can be seen from Eq.(\ref{collective}) a 
spin $I=0$ nucleus 
does not have a spectroscopic quadrupole moment $Q$ even if its
intrinsic quadrupole moment $Q_0$ is different from zero. This is 
also the case for the spin $I=1/2$ proton. 

Recently, we have determined the intrinsic quadrupole moment of the proton and 
$\Delta^+$ in the quark model~\cite{Hen01} and found 
\be
\label{intquad}
Q_0^p = -r^2_n, \qquad Q_0^{\Delta^+}  =  r^2_n, 
\ee
i.e., the intrinsic quadrupole moment of the proton is given
by the negative of the neutron charge radius and therefore
positive, whereas the intrinsic quadrupole moment of the $\Delta^+$ 
is negative. This corresponds to a prolate proton and an oblate
$\Delta^+$ deformation, consistent with the qualitative explanation 
given above.

\begin{figure}[htb]
\includegraphics{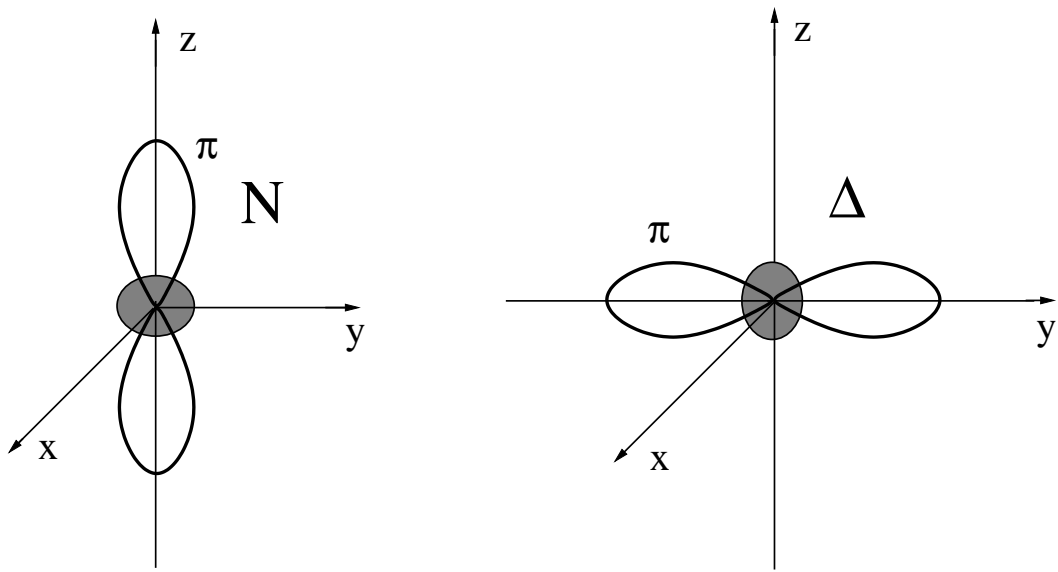}
\caption[Pion cloud ]{\label{fig:pcm}
Intrinsic quadrupole deformation of the nucleon (left)
and $\Delta$ (right) in the pion cloud model. In the $N$  
the $p$-wave pion cloud is concentrated along the polar (symmetry) axis, 
with maximum probability of finding the pion at the poles. 
This leads to a prolate deformation. In the $\Delta$, the pion cloud is 
concentrated in the equatorial plane producing an oblate intrinsic 
deformation (from Ref.~\cite{Hen01}). }
\end{figure}

The same result, namely a connection between the neutron charge radius 
$r_n^2$ and the intrinsic quadrupole moment of the proton $Q_0^p$ 
is also obtained in the pion cloud model (see Fig.~\ref{fig:pcm}). 
In the pion cloud model, the nucleon consists of a spherically symmetric 
bare nucleon (quark core) surrounded by a pion with orbital angular 
momentum $l=1$. For example, the neutron can be viewed as being composed 
of a bare proton surrounded by a negative pion. The nonspherical pion
wave function leads to prolate intrinsic deformation of the 
nucleon. For further details see Ref.~\cite{Hen01}.

Using for the nucleon the simple model of a homogeneously 
charged rotational ellipsoid, we can give $Q_0$ 
a geometric interpretation in terms of the half-axes of the ellipsoid. 
In classical electrodynamics the 
quadrupole moment of a rotational ellipsoid with charge $Z$, 
major axis $a$ along, and minor 
axis $b$  perpendicular to the symmetry axis is given by 
\begin{equation}
\label{ellipsoid}
Q_0= {2 Z \over 5} (a^2-b^2) ={4 \over 5}\,  Z \, R^2 \, \delta,
\end{equation} 
with the deformation parameter $\delta=2(a-b)/(a+b)$ and 
the mean radius $R=(a+b)/2$.  Inserting the quark model result 
$Q_0^p=-r_n^2=0.113$ fm$^2$ on the left-hand side of Eq.(\ref{ellipsoid}), 
and for $R$ the equivalent radius $R=\sqrt{5/3} \, r_p$,
where $r_p$ is the proton charge radius, we obtain 
a nucleon deformation parameter $\delta_N= 0.11 $. 
This corresponds to  a ratio of  major to minor semi-axes $a/b= 1.11$. 
For the deformation parameter of the $\Delta$ we 
find $\delta_{\Delta}=-0.09 $ and a half-axis ratio $a/b= 0.91$.

\subsection{Intrinsic quadrupole charge density of the nucleon}
\label{sec:intrinsicdensity}

To proceed, we decompose the proton charge form factor in two terms,
a term resulting from a spherically symmetric charge distribution, 
and a second term due to the intrinsic quadrupole deformation of the 
actual charge density 
\be
\label{voldefdecomp}
G_C^p(Q^2) = G_{vol}^p(Q^2) -\frac{1}{6} \, Q^2 \, G_{def}^p(Q^2).
\ee
For the spherically symmetric part we take the usual dipole 
form factor $G_D(Q^2)$. Concerning the intrinsic quadrupole charge density, 
we employ the relation between the $N \to \Delta$ charge quadrupole 
transition form factor and the elastic neutron charge form factor 
in Eq.(\ref{ffrel2}), and the relation between the intrinsic 
quadrupole moment of the nucleon and the neutron charge radius 
Eq.(\ref{intquad}). We then obtain the following expressions
\bea
\label{intrinsicC2ff}
G_{vol}^p(Q^2) & = & G_D(Q^2) = (1 + Q^2/\Lambda^2)^{-2}, 
\quad 
G_{def}^p(Q^2)  =  -\sqrt{2} \, G_{C2}^{N \to \Delta}(Q^2) 
= \frac{6}{Q^2} \, G_{C}^n(Q^2)
\nonumber \\ 
G_{C}^p(Q^2) & = & G_{vol}^p(Q^2) - G_{C}^n(Q^2), 
\eea
with $G_{def}^p(0)=Q_0^p$. Thus, the deviation of the neutron charge 
form factor from zero and the deviation of the nucleon's charge 
distribution from spherical symmetry are closely related. 

\begin{figure}[t]
\centering{
\includegraphics{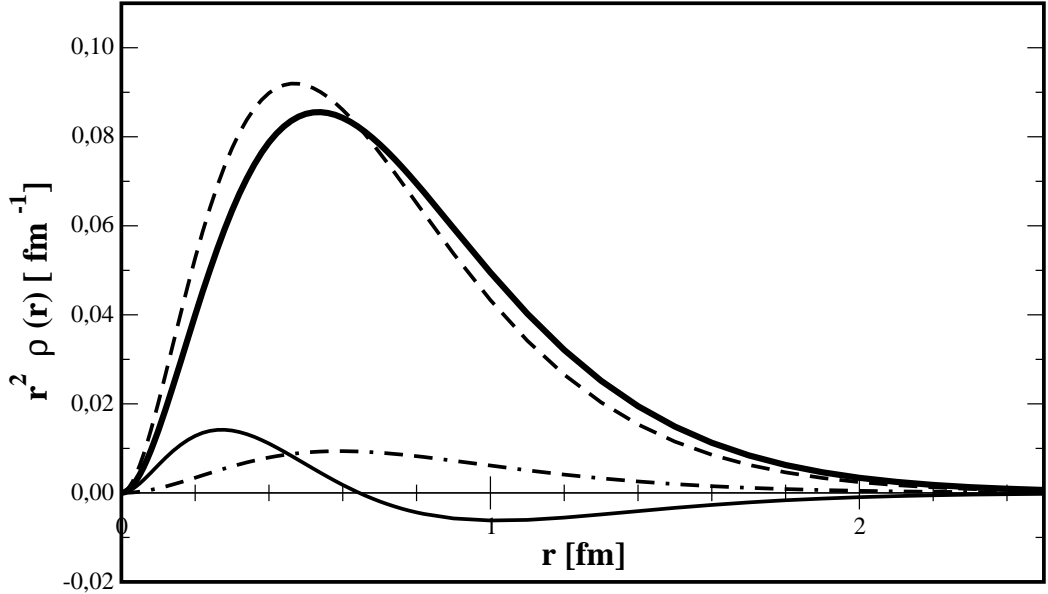}}
\caption{\label{figure:rhon} 
Spherically symmetric charge density of the proton 
$\rho_{vol}^p(r)$ [fm$^{-3}$] (broken curve);
intrinsic quadrupole charge density of the nucleon 
$\rho_{def}^p(r)$  [fm$^{-1}$] (broken-dotted curve); neutron charge density 
$\rho^n(r) = -\frac{1}{6} \, \nabla^2 \rho_{def}^p(r)$ (thin full curve); 
 and the sum $\rho^p(r) = \rho_{vol}^p - \rho^n(r)$ (thick full curve).
As a result of the intrinsic quadrupole deformation of the nucleon
positive charge is moved to the exterior region  of the proton. This will
lead to a larger proton charge radius and will be discussed in 
sect.~\ref{sec:estimate}.}
\end{figure}

According to Eq.(\ref{voldefdecomp}) the proton charge density in 
coordinate space is decomposed in two parts, a spherically 
symmetric volume term  $\rho_{vol}^p(r)$, and a nonspherical 
charge density denoted as $ \rho_{def}^p(r)$
\bea
\label{decomposition}
\rho^p(r) & = & \rho_{vol}^p(r) 
+ \frac{1}{6} \nabla^2 \rho_{def}^p(r) =  \rho_{vol}^p(r) - \rho^n(r),
\eea
where the last equality follows from Eq.(\ref{intrinsicC2ff}).
These coordinate space expressions show that the deviation of the proton 
charge density from spherical symmetry is described 
by the neutron charge density $\rho^n(r)$. 
To see the effect of the nucleon's nonsphericity explicity, 
we use a two parameter fit~\cite{Gal71} for the elastic neutron 
charge form factor
\be 
\label{Galster}
G_C^n(Q^2)  =  -\mu_n \, \frac{ a \tau}{1 + d \tau} \, G_D(Q^2), 
\ee
where $\tau = Q^2/(4 \, M_N^2)$ and $\Lambda^2=0.71$ GeV$^2$.
We then obtain for the Fourier transform of Eq.(\ref{Galster}) 
\be
\rho^n(r) = 
\frac{1}{6} r_n^2 \,
\frac{\Lambda^4 m^2}{4 \pi \, (\Lambda^2-m^2)^2} \,  
\biggl ( m^2 \frac{e^{- m r}}{r} - \Lambda^2 \frac{e^{- \Lambda r}}{r} -
(m^2-\Lambda^2)(\frac{e^{- \Lambda r}}{r} -\frac{\Lambda}{2}e^{-\Lambda r})  
\biggr ),
\ee
where $m =: 2 M_N/\sqrt{d}$.  The neutron structure parameters $a$ and $d$ 
have been determined from the lowest moments  of the experimental neutron 
charge form factor. They are given by the neutron charge radius $r_n^2$, and 
the fourth moment $r_n^4$~\cite{Gra01}. 
Similarly, the Fourier transform of the intrinsic charge quadrupole 
form factor $G_{def}^p(Q^2)$ yields
\be
\rho_{def}^p(r) = -r_n^2 \,
\frac{\Lambda^4 m^2}{4 \pi \, (\Lambda^2-m^2)^2} \,  
\biggl (\frac{e^{- m r}}{r} - \frac{e^{- \Lambda r}}{r} -
\frac{m^2-\Lambda^2}{2 \Lambda} \, e^{- \Lambda r}
\biggr ),
\ee
with $\int \! \! d^3 r \, \rho_{def}^p(r) = Q_0^p=-r_n^2$.  
Finally, the spherically symmetric part of the 
proton charge density is obtained from the Fourier transform of the 
dipole form factor 
\be 
\rho_{vol}^p(r) 
= \frac{\Lambda^3}{8 \pi} \, e^{-\Lambda r}. 
\ee
The decompostion of the proton charge density in a spherical volume 
part and a quadrupole deformation part is shown in Fig.\ref{figure:rhon}.

\section{Atomic spectroscopy and nucleon structure}

Before discussing the possibility of measuring the intrinsic
quadrupole moment of the nucleon in an atomic physics experiment,
I would like to recall that atomic spectroscopy  
provided the first clear evidence for nuclear quadrupole moments before 
these were seen with nuclear spectroscopic methods and nuclear scattering 
experiments. The quadrupole moment of the hydrogen 
isotope $D$ was discovered using the molecular beam magnetic 
resonance method~\cite{Ram39}. 

\subsection{Discovery of nuclear quadrupole moments in atomic hyperfine 
spectra}

Early in 1935, when measuring the magnetic hyperfine structure of 
certain atomic spectral lines of europium isotopes 
$^{151}{\rm Eu}$ and $^{153}{\rm Eu}$ with nuclear  spin $I=5/2$ 
Sch{\"u}ler and Schmidt found small but systematic 
deviations from the Land$\acute {\rm e}$ interval rule~\cite{Bri85}. 
The latter describes the magnetic 
hyperfine splitting of spectral lines due to the magnetic interaction energy
between a nucleus with magnetic moment $\mu_I$ and spin ${\bf I}$,  
and the magnetic moment of the atomic electrons of 
total angular momentum ${\bf J}$ 
\be
\label{magnetichfs}
\Delta W_{{hfs}} = A <{\bf I  }\cdot {\bf J}> = \frac{1}{2} \, A \, C, \qquad
A = \mu_I H(0)/(I J). 
\ee
Here,  $A$ is a constant proportional to the magnetic field 
 $H(0)$  produced by the atomic electrons at 
the site of the nucleus. Furthermore, $C=F (F+1) - I (I+1) - J (J+1)$ is 
the Casimir factor where ${\bf F} = {\bf I } + {\bf J}$ is the sum of the
nuclear and electronic angular momenta.

Sch\"uler and Schmidt noted that their experimental results could be explained 
if a deviation of the nuclear charge distribution from spherical 
symmetry is assumed. The correct quantum mechanical interpretation was given 
by Casimir~\cite{Cas36} who showed that the observed hyperfine splitting
pattern could be reproduced by adding a second term on the right hand side
of Eq.(\ref{magnetichfs}) of the form
\be
\label{chargehfs}
\Delta W_{{hfs}} = \frac{1}{2} \, A \, C + \frac{1}{8} \, B \,
\frac{3C(C+1) -4I(I+1)J(J+1)}{I(2I-1)J(2J-1)}, \qquad
B=Q \,\frac{\partial^2 \Phi}{\partial z^2},  
\ee
where the quadrupole constant $B$ is given by the product of the nuclear
quadrupole moment $Q$ and the electric field gradient 
$\frac{\partial^2 \Phi}{\partial z^2}$ due to the electrons at the 
site of the nucleus. The quadrupole moment in Eq.(\ref{chargehfs}) is 
refered to as the spectroscopic quadrupole moment. From the measured 
hyperfine splitting $\Delta W_{{hfs}}$ and the calculated electric field 
gradient the quadrupole moment $Q$ could be determined. 
For the europium isotopes the result was $Q(^{151}{\rm Eu})=150$ fm$^2$ and
$Q(^{153}{\rm Eu})=320$ fm$^2$.

For $I=0$ and $I=1/2$, Eq.(\ref{collective}) predicts $Q=0$ and a vanishing
quadrupole contribution to the hyperfine splitting $\Delta W_{{hfs}}$ 
even if the intrinsic quadrupole moment $Q_0\ne 0$.
Because of these selection rules it seems at first
sight hopeless to obtain information on the shape of an $I=1/2$ nucleus,
such as the nucleon, from an atomic physics experiment. However, there are
other observables that are sensitive to the geometric shape of the nucleus.

\subsection{Isotope shifts and intrinsic quadrupole moments}

The frequency of a certain spectral line of a given element 
is slightly different for the various isotopes of this element. 
These isotope shifts arise even in the absence of nuclear electromagnetic 
moments because different isotopes have different
mass and size.  Heavier isotopes with bigger radii experience less 
Coulomb binding than lighter isotopes with smaller radii, because
the attractive inner atomic Coulomb potential is cut off near the surface of 
nucleus. 
Consequently the spectral lines 
of the heavier isotope are slightly shifted to lower frequencies compared 
to the same lines of the lighter isotope. For $s$ electrons, the 
isotope shift is given by the following expression 
\bea
\label{isotopeshift}
\delta E_{IS} & = & \frac{2\pi}{3} \vert \Psi_e (0) \vert^2 
\delta <r^2> +  \delta E_{mass} + \delta E_{pol}, \nonumber \\
\delta <r^2>  & = &  \delta <r^2>_{vol} + \delta <r^2>_{def}
\eea
where $\delta <r^2>$ is the 
change in the nuclear charge radius of the two isotopes considered.
Here, $\delta <r^2>_{vol}$ and $\delta <r^2>_{def}$ 
denote the volume
and deformation contribution to the total charge radius change. 
The electronic wave function 
at $r=0$ is denoted as $\Psi_e(0)= \frac{1}{\pi}[Z/(na_0)]^3$, 
and $a_0$ is the Bohr radius. 
The terms $\delta E_{mass}$ and $\delta E_{pol}$ are the mass and
nuclear polarization contributions to the isotope shift. 
In the following we discuss only the size (volume) and 
shape (deformation) contributions to the isotope shift.

If the radius change in Eq.(\ref{isotopeshift}) 
were a pure volume effect, we would expect the radius variation to be 
based on $R = R_0 A^{1/3}$, where $A$ is the number of nucleons 
and $R_0 = 1.2 $ fm.  
This relation between nuclear mass and equivalent radius 
$R=\sqrt{\frac{5}{3}<r^2>}$ follows from the nuclear 
liquid droplet model, which considers a nucleus as a spherical droplet 
of constant density. On the other hand, the deformation contribution 
to the charge radius change is related to the difference of intrinsic 
quadrupole moments $Q_0$ of the considered isotopes. Thus, there can be a  
substantial radius increase if the charge distribution of one isotope is more 
deformed than the other even if their volumes are nearly the same. 

Already in 1934 Sch\"uler and Schmidt~\cite{Sch34} found an anomalously 
large isotope shift between the spectral lines of the spin $I=0$ nuclei  
 $^{150}{\rm Sm}$ and $^{152}{\rm Sm}$, which was nearly
twice as large as between neighboring isostope pairs, of mass number 
$148-150$ and
$152-154$. This large shift could not be explained 
by a pure volume effect~\footnote{A pure volume effect would give
a radius change of $\delta <r^2>=0.21$ fm$^2$, whereas the observed value
was approximately $\delta <r^2>=0.48$ fm$^2$.}. 
The explanation offered by Brix and Kopfermann in 1947~\cite{Bri49} 
is based on the following idea: (i) the anomalous isotope shift is connected  
with the jump of the quadrupole moments between 
$^{151}{\rm Eu}$ and $^{153}{\rm Eu}$, and (ii) 
the $I=0$ nuclei $^{150}{\rm Sm}$ and $^{152}{\rm Sm}$ have  
approximately the same intrinsic deformation as the $^{151}{\rm Eu}$ and
$^{153}{\rm Eu}$ nuclei with $I=5/2$. In other words, they assumed that 
the spectroscopic quadrupole moments observed in the europium isotopes 
were already present in the samarium isotopes. 
This seemed reasonable because the large quadrupole moments of the europium 
isotopes could not possibly be generated by the addition of a single proton 
to the corresponding samarium nuclei. With this assumption they could 
explain the observed large isotope shift between the two samarium isotopes.

\begin{figure}[tbh]
\centering{
\includegraphics{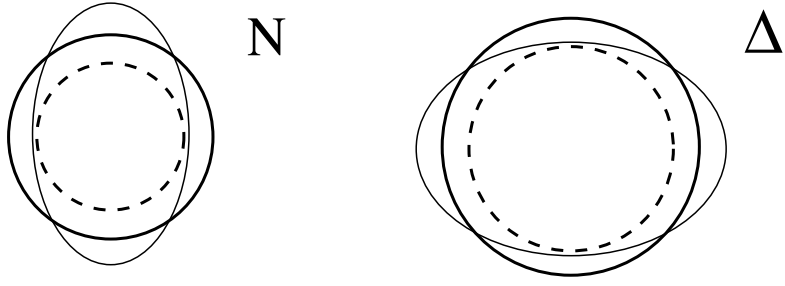}}
\caption{\label{voldef} 
Decomposition of the $N$ and $\Delta$ charge radii into a volume and a 
deformation contribution. The intrinsic quadrupole deformation of the 
nucleon and $\Delta$ leads 
to an increase of the proton (left) and $\Delta^+$ (right) charge radii. 
The broken lines correspond to the spherical volume contribution.
The deformation contribution to the 
charge radii indicated by the thin ellipsoids 
is given by the negative of the neutron charge radius according 
to Eq.(\ref{deformcont}). The full circles
represent the measured charge radii which contain the effect 
of averaging the deformed charge distribution over all directions
in space. }
\end{figure}

\subsection{Estimate of the 
nucleon deformation contribution to hydrogen hyperfine structure}
\label{sec:estimate}

Applying these ideas to the nucleon, we decompose the proton and $\Delta^+$
charge radii in two terms, a spherically symmetric 
volume contribution and a deformation contribution due to the intrinsic
quadrupole charge density (see Fig.~\ref{voldef}). 
This corresponds to the decomposition 
in Eq.(\ref{decomposition}).  The deformation contribution 
to the proton and $\Delta^+$ charge radii, which makes the charge radius
bigger, 
is determined by the neutron charge radius 
\bea 
\label{deformcont}
r^2_p & = & r^2_{p, \, vol} + r^2_{p, \, def}, \qquad 
r^2_{\Delta^+}  =   r^2_{\Delta^+, \,vol} + r^2_{\Delta^+, \, def} 
\nonumber \\ 
r^2_p& = & r^2_p + r^2_n + (- r^2_n), \qquad r^2_{\Delta^+}  =   
r^2_p + (- r^2_n). 
\eea

If written in this way we see that the volume contribution to the proton 
charge radius is given by the isoscalar charge radius 
$r^2_{p, \, vol}= r^2_{IS}$. The latter is 
defined in terms of the proton and neutron charge radii 
as $r^2_{IS}=:r^2_p+r^2_n$. The deformation contribution
to the proton charge radius is given by the negative of the neutron charge
radius $r^2_{p, \, def}= -r^2_{n}$. Similarly, the volume contribution to
the $\Delta^+$ charge radius is given by $r^2_{\Delta^+, \, vol}= r^2_p$,
and the deformation contribution is (as in the case of the proton) given by
$r^2_{\Delta^+, \, def}= -r^2_n$. 

In the quark model the isoscalar charge radius is mainly determined by 
the quark core radius $b \approx 0.6$ fm and the charge radius of the 
constituent quark $r_q \approx 0.6 $ fm 
\be 
r^2_{IS} = b^2 + r^2_q. 
\ee
The quark core of the nucleon and the constituent quarks 
themselves are assumed to be spherical. Concerning the quark core, this is 
supported by explict calculation of the $P$ and 
$D$ wave probabilities in the nucleon wave function, which are 
very small~\cite{Buc91}.
The deformation of the nucleon arises predominantly 
from the collective quark-antiquark degrees of freedom described 
by the two-body operators of section~\ref{sec:NDelta}.

To estimate the energy shift of $s$ electrons in atomic hydrogen 
which results from the deformation of the nucleon's charge density 
we use  $\delta <r^2>= - r_n^2$ in Eq.(\ref{isotopeshift}). 
For $n=1$ and $a_0=0.529 $ \AA \ we then obtain 
\be 
\delta E_{IS}({\rm def})= 157 \, {\rm kHz}. 
\ee
Compared to other corrections in the hydrogen hyperfine 
structure~\cite{Dup03} this seems to be a big effect. 
We emphasize that this deformation contribution is already included in 
the experimental proton charge radius~\cite{Ude97}. 
One possibility to isolate the deformation contribution to the proton 
charge radius is to measure hydrogen isotope shifts. 
The increasing accuracy of such 
experiments~\cite{Hub97} will perhaps reveal some 
discrepancies between theory and experiment that arise from 
the intrinsic deformation of the hydrogen isotopes,
and which manifest themselves as deformation contributions to their 
charge radii.
In this connection we mention that the proton charge radius as determined 
in hydrogen Lamb shift measurements may be principally different 
from the one determined in elastic electron-proton scattering.
Atomic physics experiments involve small electronic line widths 
and thus measure the electromagnetic interaction of the bound electron 
with the proton over long time scales. Thus they see a time average of 
the proton's deformed  charge distribution. Because of this averaging effect 
they yield a bigger proton charge radius. On the other hand, high energy 
electrons involve comparatively short interaction times, 
which corresponds to taking a snapshot of the deformed charge distribution 
having a particular orientation in space and thus see a slightly
smaller charge radius.

\section{Summary}

In summary, the electromagnetic excitation of the nucleon to
its first excited state, the $\Delta$ resonance, has provided 
clear evidence that the charge distribution of the nucleon 
deviates from spherical symmetry. The intrinsic quadrupole moment 
of the nucleon, which is defined with respect to a body-fixed coordinate 
system that co-rotates with the nucleon, 
is a measure of the nucleon's quadrupole deformation.
We have calculated the intrinsic quadrupole 
moment of the nucleon in the quark model and found that it is given by the 
neutron charge radius, implying a prolate deformation of the nucleon's 
charge distribution. More generally, we have suggested that the neutron 
charge density $\rho^n(r)$ is a measure of the 
intrinsic quadrupole charge density of the nucleon $\rho^p_{def}(r)$. 

Isotope shifts of atomic spectral lines have provided 
information on the intrinsic deformation of spin 1/2 and spin 0 nuclei,
which do not have  spectroscopic quadrupole moments due to angular momentum
selection rules. The hydrogen spectrum can be measured with very high 
accuracy. It does not seem completely unlikely that 
some future experimental technique, perhaps involving muonic hydrogen,
or hydrogen molecules, such as $H_2$, $HD$, and $D_2$, will reveal further 
nucleon structure details such as the spatial shape of the proton.

\end{document}